\documentclass[12pt]{article}
\usepackage{a4wide,epsfig,psfrag,amsmath,amssymb,cite,scalefnt}
\usepackage{color}

\newcommand{\HPL}[2]{{\rm H}_{#1}(#2)}
\newcommand{\HS}[2]{{\rm S}_{#1}(#2)}

\newcommand{\gsim}{\;\rlap{\lower 3.5 pt \hbox{$\mathchar \sim$}} \raise 1pt
 \hbox {$>$}\;}
\newcommand{\lsim}{\;\rlap{\lower 3.5 pt \hbox{$\mathchar \sim$}} \raise 1pt
 \hbox {$<$}\;}

\allowdisplaybreaks

\begin{document}

\title{\vskip-3cm{\baselineskip14pt
    \begin{flushleft}
      \normalsize SFB/CPP-12-93\\
      \normalsize TTP12-45\\
      \normalsize LPN12-127
  \end{flushleft}}
  \vskip1.5cm
  Higgs boson production at the LHC: NNLO partonic cross sections through
  order $\epsilon$ and convolutions with splitting functions to 
  N$^3$LO
}

\author{
  Maik H\"oschele, Jens Hoff, Alexey Pak,
  \\
  Matthias Steinhauser, Takahiro Ueda
  \\[1em]
  {\small\it Institut f{\"u}r Theoretische Teilchenphysik}\\
  {\small\it Karlsruhe Institute of Technology (KIT)}\\
  {\small\it 76128 Karlsruhe, Germany}
}

\date{}

\maketitle

\thispagestyle{empty}

\begin{abstract}
  We consider Higgs boson production at hadron colliders 
  in the gluon fusion channel and compute higher order terms in the
  regularization parameter $\epsilon$. In particular, 
  the next-to-next-to-leading order
  cross section is evaluated including order $\epsilon$ terms. 
  These results are used 
  to compute all convolutions with the splitting functions 
  entering the next-to-next-to-next-to-leading order cross section.
  \medskip

\end{abstract}

\thispagestyle{empty}

%- }}}

\newpage

%- {{{ Introduction:

\section{Introduction}

After the recent discovery of a new boson at the Large Hadron Collider (LHC)
at CERN~\cite{:2012gk,:2012gu} one has to clarify the
question whether it is the Standard Model (SM) Higgs boson or a new particle
of an enlarged theory.  For this purpose it is important to study in detail
its production and decay processes. In the SM, but also in many extensions,
the largest 
production cross section for a Higgs boson is given by the gluon fusion
sub-process which, however, also has a large uncertainty. The
uncertainties
arise mainly from the parton distribution functions (PDFs) and
from unknown higher order corrections. The latter amount to about 10\% 
although
perturbative next-to-next-to-leading order 
(NNLO) QCD and NLO electroweak corrections
have been computed in the recent years and are thus available for theory
predictions (see Refs.~\cite{Dittmaier:2011ti,Dittmaier:2012vm} for
comprehensive reviews).  In view of the expected experimental precision
reached by the LHC experiments it is desirable to take the next step and
compute the N$^3$LO QCD corrections to the gluon fusion process.  
In this letter
we contribute important building blocks to this enterprise by computing the
lower order partonic cross sections to higher order in $\epsilon=(4-D)/2$,
where $D$ is the space-time dimension, and by
providing results for their convolution with the splitting functions.
In our calculation we will make use of the results for the master integrals
entering the NNLO calculation. In Refs.~\cite{Pak:2011hs,Anastasiou:2012kq}
results for the $\epsilon$ expansion are provided which are sufficient for
the N$^3$LO predictions.

The starting point for the evaluation of the cross section is the
effective five-flavour Lagrange density
\begin{eqnarray}
  {\cal L}_{Y,\rm eff} &=& -\frac{H^0}{v^0} C_1^0 {\cal O}_1^0 
  + {\cal L}_{QCD}^{(5)}
  \,,
  \label{eq::leff}
\end{eqnarray}
where ${\cal L}_{QCD}^{(5)}$ is the usual QCD part with five massless quarks,
$H^0$ denotes the Higgs boson, $v^0$ its vacuum expectation value and $C_1^0$
is the matching coefficient between the full and the effective theory.
$G_{\mu\nu}^0$ is the gluonic
field strength tensor constructed from fields and couplings already present in
${\cal L}_{QCD}^{(5)}$. The superscript ``0'' denotes the bare quantities. Note
that the renormalization of $H^0/v^0$ is of higher order in the
electromagnetic coupling constant.  In Eq.~(\ref{eq::leff}) the operator
${\cal O}_1^0$ is given by
\begin{eqnarray}
  {\cal O}_1^0 &=& \frac{1}{4} G_{\mu\nu}^0 G^{0,\mu\nu}\,.
\end{eqnarray}
As shown in
Refs.~\cite{Harlander:2009bw,Pak:2009bx,Harlander:2009mq,Pak:2009dg,Harlander:2009my,Pak:2011hs},
through NNLO the results obtained in the effective-theory approach agree to
good accuracy with the ones of the full-theory calculation.
We assume that this feature also holds at N$^3$LO.

The outline of the paper is as follows: in the next two Sections we discuss
in detail the splitting functions and the partonic cross sections up to NNLO.
The procedure to compute the convolution integrals is described in
Section~\ref{sec::conv} where also some sample results are given.
Analytical results for all convolutions are provided in a 
{\tt Mathematica} file which can be downloaded from~\cite{progdata}.
In Section~\ref{sec::sum} we summarize our findings.

%- }}}
%- {{{ Splitting functions:

\section{Splitting functions}

In general, the sum of the renormalized real and the virtual contributions
of a given partonic sub-process is not finite in the limit
$\epsilon\to 0$. The remaining poles originate from the collinear divergences
in the initial state. For example, the initial gluon may split into a
quark-antiquark 
pair and the quark participates in the Higgs boson production.  It also may
happen that a quark in the initial state radiates a gluon before participating
in the hard scattering process.

The correct treatment of the singularities is achieved by the
convolution of the partonic cross section with the splitting functions
$P_{ij}(x)$ describing the probability of parton $j$ to emit a parton $i$ 
with the fraction $x$ of its initial energy. The
perturbative expansion is given by
\begin{eqnarray}
  P_{ij}(x) &=& \delta_{ij} \delta(1 - x)
  + \frac{\alpha_s^{(5)}}{\pi} P_{ij}^{(1)}(x) 
  + \left(\frac{\alpha_s^{(5)}}{\pi}\right)^2 P_{ij}^{(2)}(x) 
  + \left(\frac{\alpha_s^{(5)}}{\pi}\right)^3 P_{ij}^{(3)}(x) 
  + \ldots\,,
  \label{eq::Pij}
\end{eqnarray}
where the analytical results for $P_{ij}^{(k)}(x)$ can be found in
Refs.~\cite{Altarelli:1977zs,Curci:1980uw,Moch:2004pa,Vogt:2004mw}.
It is common practice to refer to $P_{ij}^{(k)}(x)$ as the $k$-loop 
splitting function.

We have written the expansion in Eq.~(\ref{eq::Pij}) in terms of
$\alpha_s^{(5)}$, the renormalized coupling
in the effective theory with decoupled top quark and $n_l = 5$
massless quarks.

In our calculation we use the splitting functions defined in the
$\overline{\rm MS}$ scheme. This is consistent with the definition of the
parton distribution functions (PDFs) which absorb the non-perturbative and
non-singular features of the initial state.

For definiteness let us present the explicit results for the one-loop
splitting functions which read\footnote{Note that our definition of
  $P^{(k)}_{qg}$ differs by a factor $2n_l$ from that of
  Ref.~\cite{Vogt:2004mw}.} 
\begin{eqnarray}
  P^{(1)}_{qq} &=& -\frac{2}{3} - \frac{2x}{3} + \delta(1-x)
  + \frac{4}{3}\left[\frac{1}{1-x}\right]_+ \,,\nonumber\\
  P^{(1)}_{qg} &=& \frac{1}{4}- \frac{x}{2} + \frac{x^2}{2} \,,\nonumber\\
  P^{(1)}_{gq} &=& -\frac{4}{3} + \frac{4}{3x} + \frac{2x}{3} \,,\nonumber\\
  P^{(1)}_{gg} &=& -6 + \frac{3}{x} + 3x - 3x^2 
  + \left(\frac{11}{4} - \frac{n_l}{6} \right) \delta(1-x)
  + 3\left[\frac{1}{1-x}\right]_+\,,
\end{eqnarray}
where $n_l$ is the number of active flavours.
As far as the two-loop splitting function $P^{(2)}_{qq}$ is concerned we use
$P_{qq} = P_{\rm ns}^+ + P_{\rm ps}$~\cite{Vogt:2004mw} where 
the two-loop terms of $P_{\rm ns}^+$ and $P_{\rm ps}$ are given in
Eq.~(4.6) of Ref.~\cite{Moch:2004pa} and Eq.~(4.7) of Ref.~\cite{Vogt:2004mw}, 
respectively. The analytical result reads
\begin{eqnarray}
  P^{(2)}_{qq} &=& 
  \frac{1}{2} - \frac{143x}{18} 
  + \pi^2\left( \frac{4}{27} + \frac{5x}{27} + \frac{1}{27(1 + x)} \right)
  + n_l \left(-\frac{19}{27} + \frac{20}{27x} + \frac{65x}{27} -
    \frac{56x^2}{27}
  \right)
  \nonumber \\&&\mbox{}
  + \left[-\frac{17}{18} + \frac{7}{3(1 - x)} 
    - \frac{11x}{6} + n_l\left(\frac{4}{9} - \frac{2}{9(1 - x)} + \frac{16x}{9} 
      + \frac{8x^2}{9} \right)
    \right] \HPL{0}{x}
  \nonumber \\&&\mbox{}
  + \left(-\frac{8}{9} + \frac{16}{9(1 - x)} - \frac{8x}{9} 
    \right)\HPL{2}{x}
  + \left(-\frac{2}{9} + \frac{2x}{9} + \frac{4}{9(1 + x)}\right)\HPL{-1, 0}{x}
  \nonumber \\&&\mbox{}
  + \left[-\frac{4}{3} + \frac{2}{(1 - x)} 
    - \frac{14x}{9} - \frac{2}{9(1 + x)}
    + n_l\left(-\frac{2}{3} - \frac{2x}{3}\right)
  \right]\HPL{0, 0}{x}
  \nonumber \\&&\mbox{}
  + \left(-\frac{8}{9} + \frac{16}{9(1 - x)} - \frac{8x}{9}\right)\HPL{1, 0}{x}
  + \left(\frac{67}{9} -\frac{\pi^2}{3} - n_l\frac{10}{27}\right)
  \left[\frac{1}{1-x}\right]_+
  \nonumber \\&&\mbox{}
  + \left[\frac{7}{8} 
    + \frac{7\pi^2}{18} - \frac{\zeta(3)}{3}
    + n_l\left(-\frac{1}{36} - \frac{\pi^2}{27}\right) 
  \right]\delta(1-x)
  \,,
\end{eqnarray}
where $\zeta(n)$ is Riemann's zeta function.
The remaining two-loop splitting functions can be found in
Eqs.~(4.8),~(4.9) and~(4.10) and the three-loop ones in
Eqs.~(4.14) and~(4.15) of Ref.~\cite{Vogt:2004mw}.
Note that at the three-loop level only $P^{(3)}_{gq}$ and
$P^{(3)}_{gg}$ are needed.

%- }}}
%- {{{ Partonic cross sections:

\section{Partonic cross sections}

The renormalized partonic cross section at N$^3$LO receives 
contributions from
convolutions of splitting functions with
the renormalized lower order partonic cross sections.  For
this purpose the latter needs to be expanded to higher order in $\epsilon$
since the convolutions with $P_{ij}^{(k)}(x)$ come along with poles in
$\epsilon$.\footnote{We refer to~\cite{Anastasiou:2002yz,Ravindran:2003um} for
  a nice description of the procedure in case of NNLO Higgs boson
  production. In particular, in Ref.~\cite{Anastasiou:2002yz} one finds
  explicit formulae showing how the convolutions of splitting functions and
  partonic cross sections enter the renormalized result.}  Furthermore,
$\alpha_s$ and the operator ${\cal O}_1$ have to be renormalized which means
that the LO, NLO and NNLO partonic cross sections are multiplied by
contributions containing $1/\epsilon$ poles and thus the $\epsilon$ expansion
needs to be sufficiently deep.

We have performed two independent calculations of the partonic cross
sections. On the one hand we extended the full-theory calculation of
Ref.~\cite{Pak:2009dg} to one order higher in $\epsilon$ and have computed the
leading term in the heavy-top quark mass expansion.  On the other hand we have
implemented the effective-theory Feynman rules (cf. Eq.~(\ref{eq::leff})) 
and evaluated the partonic cross sections of all sub-processes.
Of course, the NLO and the NNLO results present in the
literature~\cite{Dawson:1990zj,Djouadi:1991tka,Spira:1995rr,Harlander:2002wh,Anastasiou:2002yz,Ravindran:2003um}
could be reproduced. The ${\cal O}(\epsilon^2)$ terms at NLO and the ${\cal
  O}(\epsilon)$ terms at NNLO are new.

From Eq.~(\ref{eq::leff}) it is clear that the partonic cross sections
contain the two factors,
the matching coefficient $C_1$ and the cross section
computed using the effective Higgs-gluon coupling contained in
$H^0 {\cal O}_1^0$. Thus, for the renormalized cross section we can write
\begin{eqnarray}
  \hat\sigma_{ij} &=& C_1^2 \tilde\sigma_{ij}\,,
\end{eqnarray}
with $C_1 = - \alpha_s^{(5)}/(3\pi)(1+{\cal O}(\alpha_s))$.
In the remainder of the paper we will only consider $\tilde\sigma_{ij}$, which
we refer to as the ``reduced partonic cross section'', 
since only this quantity enters the convolutions with the splitting functions;
the multiplication with the finite factor $C_1^2$ can be
done at the very end.
We note that the matching coefficient $C_1$ has been computed to
three-~\cite{Chetyrkin:1997un,Kramer:1996iq} and four-loop 
order~\cite{Chetyrkin:1997un,Schroder:2005hy,Chetyrkin:2005ia},
respectively. 

In analogy to Eq.~(\ref{eq::Pij}) we write the perturbative expansion of the 
reduced partonic cross section as
\begin{eqnarray}
  \tilde\sigma_{ij}(x) &=& A 
  \Bigg[ 
  \tilde\sigma^{(0)}_{ij}(x)
  + \frac{\alpha_s^{(5)}}{\pi} \tilde\sigma^{(1)}_{ij}(x) 
  + \left(\frac{\alpha_s^{(5)}}{\pi}\right)^2 \tilde\sigma^{(2)}_{ij}(x) 
  + \ldots
  \Bigg]\,,
  \label{eq::sigtil}
\end{eqnarray}
with $A = G_F \pi /(32\sqrt{2})$ and $x=M_h^2/s$.
The dependence on $\mu^2/M_h^2$, where $\mu$ represents the renormalization
and the factorization scale, is suppressed.
At NLO we have $i,j\in\{g,q,\bar{q}\}$ and at the NNLO
the possible initial states are given by
$gg$, $qg$, $\bar{q}g$, $q\bar{q}$, $qq$, or $qq^\prime$, where $q$ and $q^\prime$
stand for (different) massless quark flavours.\footnote{It is understood
that ghosts are always considered together with gluons.}
The computation of the N$^3$LO contributions requires 
that $\tilde\sigma^{(k)}$ is known to order $\epsilon^{3-k}$.

In order to fix the notation we provide the explicit results for
$\tilde\sigma^{(0)}_{ij}$ and $\tilde\sigma^{(1)}_{q\bar{q}}$
which are given by
\begin{eqnarray}
  \tilde\sigma^{(0)}_{ij} &=& \delta_{ig}\delta_{jg}\frac{\delta(1-x)}{1-\epsilon}
  \,,
  \nonumber\\
  \tilde\sigma^{(1)}_{q\bar{q}} &=& \frac{32}{27}(1-x)^3\Bigg\{
  1 + \epsilon\left(\frac{2}{3} + \HPL{0}{x_{\mu H}} 
    - 2\HPL{0}{1-x} + \HPL{0}{x}\right)
  \nonumber\\&&\mbox{}
  + \epsilon^2\Bigg[\frac{13}{9} - \frac{\pi^2}{4} +
    \frac{1}{2}\left(\HPL{0}{x_{\mu H}}\right)^2
    + 2\left(\HPL{0}{1-x}\right)^2 
    + \HPL{0}{1 - x}\left(-\frac{4}{3} - 2\HPL{0}{x}\right) 
    \nonumber\\&&\mbox{}
    + \frac{2}{3}\HPL{0}{x} 
    + \frac{1}{2}\left(\HPL{0}{x}\right)^2 
    + \HPL{0}{x_{\mu H}}\left(\frac{2}{3} - 2\HPL{0}{1 - x} + 
      \HPL{0}{x}\right)
  \Bigg]
  \Bigg\}\,,
\end{eqnarray}
with $x_{\mu H} = \mu^2/M_H^2$.
The other results for the partonic cross sections are significantly
larger and are given in computer readable form in Ref.~\cite{progdata}.

%- }}}
%- {{{ Convolutions of partonic cross sections and splitting functions:

\section{\label{sec::conv}Convolutions of partonic cross sections and splitting
  functions}

The convolution of functions $f$ and $g$ that enter
$P_{ij}^{(k)}(x)$ and $\hat{\sigma}_{ij}$ is
defined as
\begin{eqnarray}
  \left[f \otimes g \right](x)
  &=& \int_0^1 {\rm d}x_1 {\rm d}x_2 \delta(x - x_1 x_2) f(x_1) g(x_2)
  \,.
\end{eqnarray}
Both the results for the partonic cross sections (see previous
Section) and the splitting functions include combinations of HPLs 
up to weight four with 
factors $1/x$, $1/(1-x)$, and $1/(1+x)$, and the
generalized functions $\delta(1-x)$ and
$\left[\frac{\ln^k(1-x)}{1-x}\right]_+$.  In the following we briefly describe
our approach to compute the convolutions of such functions.
A detailed description can also be found in Appendix~B of
Ref.~\cite{Pak:2011hs}.

\begin{enumerate}
\item First, we transform the convoluted expressions to Mellin space via
  \begin{eqnarray}
    M_n\left[f(x)\right] &=& \int_0^1 {\rm d}x \, x^{n-1} f(x)\,.
    \label{eqn:mellinn}
  \end{eqnarray}
  This transformation turns convolutions into products due to the identity
  \begin{eqnarray}
    M_n\Big[\left[f \otimes g\right](x)\Big] &=& M_n\left[f(x)\right]
    M_n\left[g(x)\right]\,. 
  \end{eqnarray}
  A comprehensive discussion of the Mellin transform and the list of all Mellin
  images appearing in the calculation of the NNLO Higgs boson production rate
  can be found in Refs.~\cite{Blumlein:1998if,Ablinger:2011te}. 
  Mellin transforms of HPLs and their derivatives can be conveniently expressed
  in terms of harmonic sums~\cite{Vermaseren:1998uu,Remiddi:1999ew}.
  In our calculation, we used the {\tt FORM} package {\tt harmpol}~\cite{harmpol} 
  to perform the transformation.

\item Second, we prepare a table of Mellin transforms of HPLs to certain maximum 
  weight. For example, the table through weight one reads
  \begin{eqnarray}
    M_n[1] &=& \frac{1}{n}\,,\nonumber\\\nonumber
    M_n[\HPL{0}{x}] &=& - \frac{1}{n^2}\,,\\ \nonumber
    M_n[\HPL{1}{x}] &=& \frac{\HS{1}{n}}{n}\,,\\ 
    M_n[\HPL{-1}{x}]&=& - \frac{(-1)^n}{n} \left( \HS{-1}{n} + \ln{2}
    \right) + \frac{\ln{2}}{n}\,,
    \label{eqn:mttab}
  \end{eqnarray}
  where $\HPL{i}{x}$ are the HPLs of weight one and
  $\HS{i}{n}$ are the harmonic sums of weight one.
  Mellin transforms of HPLs of higher weights produce harmonic sums of higher
  weights and transcendental numbers originating from HPLs evaluated at $x=1$.

\item Third, we prepare a table of Mellin transforms of ``regularized derivatives'' of 
  HPLs in the table above as follows:
  \begin{eqnarray}
    && M_n\left[ \hat{\partial}_x 1 \right] = 1, \\
    && M_n\left[ \hat{\partial}_x f(x) \right] = R\left[f(x)\right]
    - (n-1) M_{n-1}\left[ f(x) \right],
    \label{eq::regderiv}
  \end{eqnarray}
  where
  \begin{eqnarray}
    && R\Big[g_k(x)\ln^k(1-x) + g_{k-1}(x)\ln^{k-1}(1-x) + ... + g_0(x)\Big] = g_0(1),
    \\ \nonumber
    && \mbox{with}~~ g_{j}(1) \ne 0~~ \forall {j} > 0.
    \label{eqn:rderiv}
  \end{eqnarray}
  This definition allows us to establish relations between the
  regularized derivatives of HPLs and the ``common'' generalized functions. 
  In particular, 
  \begin{eqnarray}
    \hat{\partial}_x~ 1 &=& \delta(1 - x), \nonumber\\
    \hat{\partial}_x \HPL{1}{x} &=& \left[\frac{1}{1-x}\right]_+, \nonumber\\
    \hat{\partial}_x \HPL{11}{x} &=& -\left[\frac{\ln(1-x)}{1-x}\right]_+, \nonumber\\
    \hat{\partial}_x \HPL{111}{x} &=& \frac{1}{2}
    \left[\frac{\ln^2(1-x)}{1-x}\right]_+,
    \nonumber\\
    \hat{\partial}_x \HPL{101}{x} &=& \frac{\pi^2}{6} \left[\frac{1}{1-x}\right]_+
    + \frac{\HPL{01}{x} - \zeta_2}{1-x}\,.
  \end{eqnarray}
  Note that if $f(x)$ is not divergent at $x=1$, its regularized derivative 
  Eq.~\eqref{eq::regderiv} reduces to the usual derivative:
  \begin{equation}
    M_n\left[ \frac{d}{dx} f(x) \right]
    = x^{n-1} f(x) \Big|_0^1 - (n-1) M_{n-1} \left[ f(x) \right] .
  \end{equation}
  For example,
  \begin{align}
    \begin{split}
      \hat{\partial}_x \HPL{-1}{x} &= \frac{d}{dx} \HPL{-1}{x} = \frac{1}{1+x} , \\
      \hat{\partial}_x \HPL{-1,1}{x} &= \frac{d}{dx} \HPL{-1,1}{x} = \frac{\HPL{1}{x}}{1+x} .
    \end{split}
  \end{align}
  Thus, it is not necessary to separately consider Mellin transforms of
  all functions appearing in the derivatives of HPLs, such as
  $\frac{\HPL{...}{x}}{1+x}$,
  $\left[\frac{\HPL{...}{x}}{1-x}\right]_+$ and so on.
  Rather it is sufficient to consider Mellin transforms of regularized derivatives in
  case Mellin transforms for generalized functions are needed.
  For illustration we list the Mellin transforms of regularized derivatives
  of HPLs to weight one
  \begin{eqnarray}
    M_n[\hat{\partial}_x 1] &=& 1\,, \nonumber\\
    M_n[\hat{\partial}_x \HPL{0}{x}] &=& \frac{1}{n-1}\,, \nonumber\\
    M_n[\hat{\partial}_x \HPL{1}{x}] &=& -\HS{1}{n-1}\,, \nonumber\\
    M_n[\hat{\partial}_x \HPL{-1}{x}] &=& 
    (-1)^{n-1}\HS{-1}{n-1} + (-1)^{n-1}\ln{2}\,.
    \label{eq::MTgenderiv}
  \end{eqnarray}

\item Finally, we prepare and solve the system of linear equations and thus perform the inverse Mellin transform.
  Combining the tables for Mellin transforms of HPLs as
  in Eqs.~\eqref{eqn:mttab}
  and the regularized derivatives of HPLs as in Eqs.~\eqref{eq::MTgenderiv}
  one obtains a system of linear equations for terms of the form $1/n^k$,
  $\HS{...}{n}/n^k$ and $(-1)^n \HS{...}{n}/n^k$
  which can be used in order to find the inverse Mellin transforms. In
  principle, the  
  expression in the Mellin space may have other terms, whose inverse transform 
  cannot be determined from the system, but all such terms cancel in the real calculation. 
\end{enumerate}

The above steps have been implemented in a {\tt Mathematica} program
which uses the {\tt HPL} package~\cite{Maitre:2005uu,Maitre:2007kp}
and can compute all the necessary convolutions. 

As an important cross check,
we have performed all the convolution integrals also numerically
for some specific value of $x$. In case no plus distribution is present in the
integrand it is straightforward to perform the numerical integration.
For convolution integrals involving one or more plus distributions,
the cancellation 
of the singular behaviour among terms can be quite involved.
The problem can be solved by introducing auxiliary 
regularizations~\cite{Anastasiou:2005qj}
\begin{equation}
  \frac{\ln^n(1-x)}{1-x}
  \to
  \lim_{\eta \to 0} \lim_{a \to 1}
  \frac{1}{\eta^n}
  \frac{\partial^n}{\partial a^n}
  (1-x)^{-1+a \eta} ,
\end{equation}
for evaluating integrals of the individual terms.
After performing the integrals by using $\delta$ functions in the definition of
the convolutions one rescales the integration variables such that the integration
domain becomes the unit hypercube. The
singularities come from both the lower and the upper limit of
the integration as well as from the overall 
$(1-x)^{-1+a \eta}$ factors which give
$\delta$ functions and plus distributions in the result of the convolution,
after applying
\begin{equation}
  \lambda^{-1+a \eta}
  =
  \frac{\delta(\lambda)}{a \eta}
  + \sum_{j=0}^\infty \frac{(a \eta)^j}{j!}
    \left[ \frac{\ln^j(\lambda)}{\lambda} \right]_+ .
\end{equation}
In the next step one differentiates the result with respect to $a$, takes the
limit $a \to 1$, and finally expands in $\eta$.
All pole terms $1/\eta^k$ should cancel among the integrals and hence
the leading term gives the result of the convolution.
The remaining integrals have no divergences and can be
performed numerically.

As a further welcome check we have compared our results with the explicit
expressions for the convolutions given in the Appendix of
Ref.~\cite{Ridder:2012dg}. In particular we found complete agreement
with Eqs.~(C.28) to~(C.31) which involve convolutions of two plus
distributions.

In Tab.~\ref{tab::conv} we classify the convolutions which are required
for  
the N$^3$LO calculation according to the number of nested convolutions. 
Contributions where only one convolution is involved already enter the NLO and 
the NNLO results.
At the N$^3$LO, the LO, NLO and NNLO partonic cross section has to be convoluted
with three-, two- and one-loop splitting functions, respectively.
Two convolution integrals are only present at NNLO and N$^3$LO and three
convolutions 
only at N$^3$LO. In the latter case, however, only one-loop splitting
functions and the leading order cross section enter. Since the latter is proportional to 
$\delta(1-x)$, only two non-trivial integrals remain.

The right-most column in Tab.~\ref{tab::conv} defines the name of the file
where the analytical results can be found. At NLO there are only 2 non-zero
convolutions, at NNLO there are already 18 and at N$^3$LO altogether 82
convolutions of partonic cross sections with up to three splitting functions.
All the results can be downloaded from~\cite{progdata} where for convenience we
also provide intermediate results for the convolutions involving only
splitting functions. In Appendix~\ref{app::conv} we provide a list assigning
the convolutions to the file names. Note that not all results are shown which
are identical due to symmetry.

\begin{figure}[t]
  \centering
  \begin{tabular}{ccc}
    \includegraphics[width=0.25\linewidth]{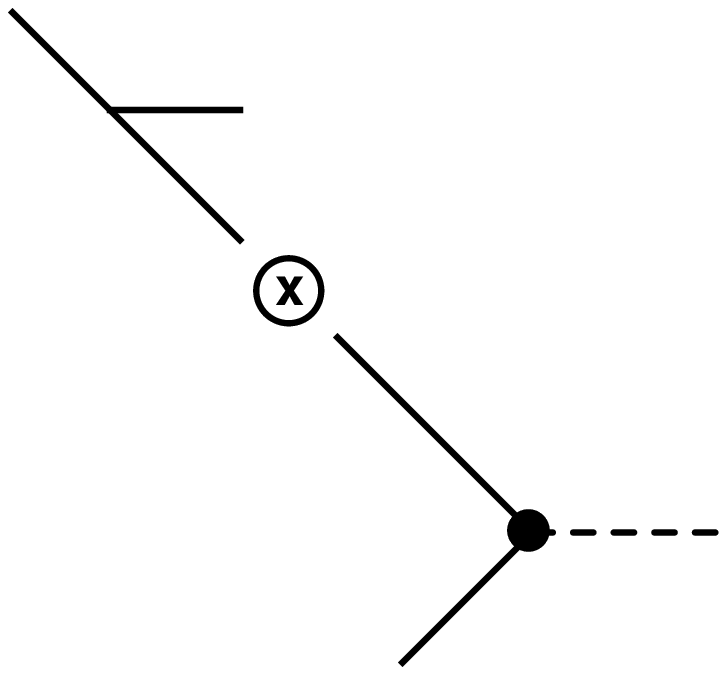}
    &
    \includegraphics[width=0.25\linewidth]{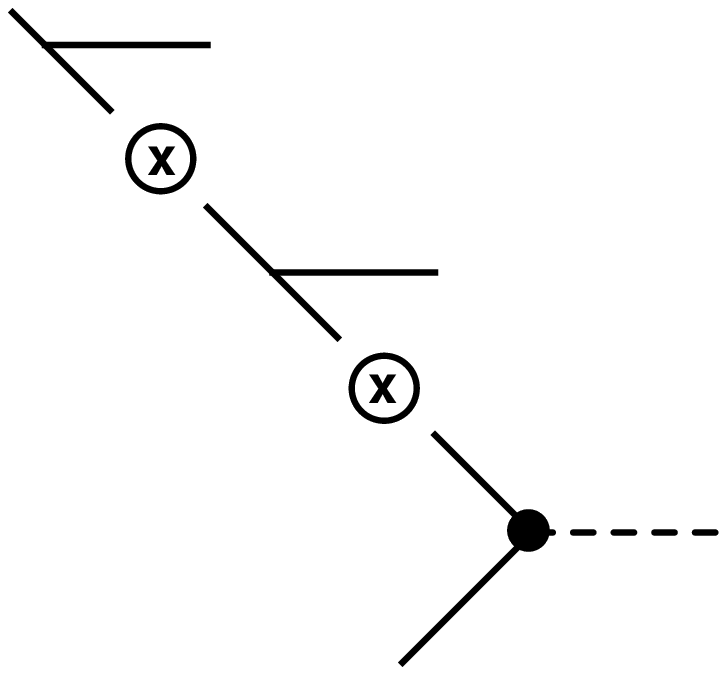}
    \qquad\qquad&
    \includegraphics[width=0.20\linewidth]{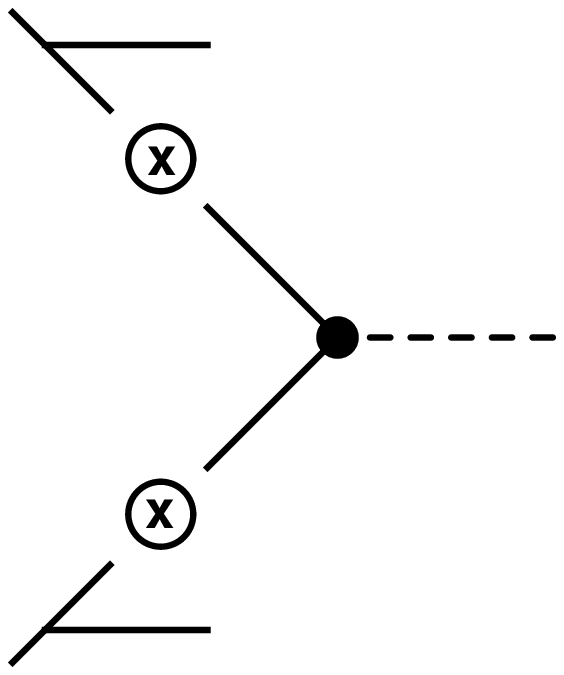}
    \\
    $(\tilde\sigma_{ij}^{(n)}/x) \otimes P_{jk}^{(r)}$ &
    $(\tilde\sigma_{ij}^{(n)}/x) \otimes P_{jk}^{(r)}\otimes P_{kl}^{(s)}$ &
    $P_{il}^{(s)} \otimes (\tilde\sigma_{ij}^{(n)}/x) \otimes P_{jk}^{(r)}$ 
    \\ [2em]
    \multicolumn{3}{c}{
      \includegraphics[width=0.25\linewidth]{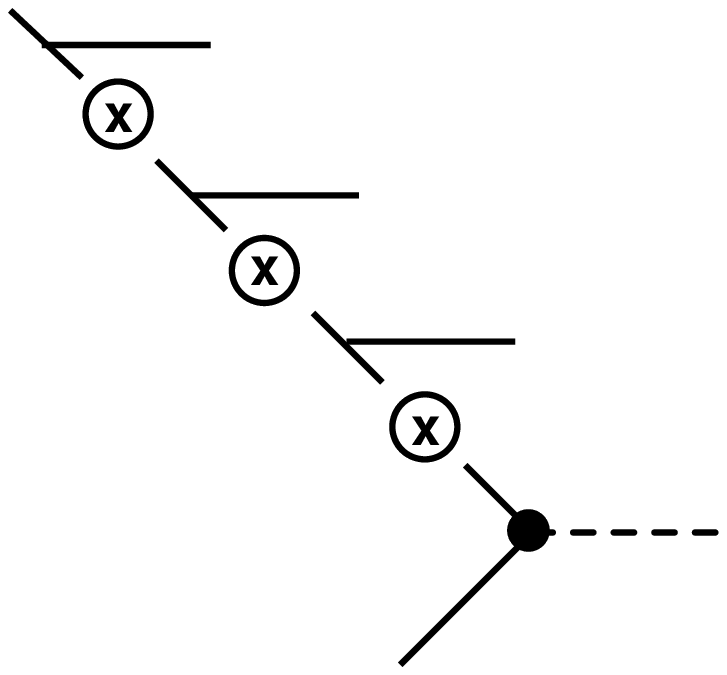}
      \qquad\qquad
      \includegraphics[width=0.22\linewidth]{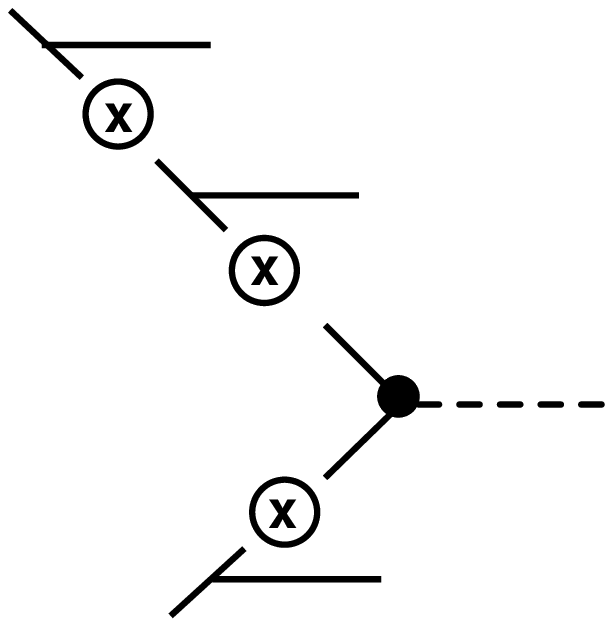}
      }
    \\
    \multicolumn{3}{c}{
      $(\tilde\sigma_{ij}^{(0)}/x) \otimes P_{jk}^{(1)}\otimes P_{kl}^{(1)} 
      \otimes P_{lm}^{(1)}$
      \qquad\qquad
      $P_{im}^{(1)} \otimes (\tilde\sigma_{ij}^{(0)} /x)
      \otimes P_{jk}^{(1)}\otimes P_{kl}^{(1)}$
    }
  \end{tabular}
  \caption[]{\label{fig::sigxp}
  Schematic representation of the convolutions of the partonic cross section
  $\tilde\sigma_{ij}$ with the splitting functions $P_{jk}$.}
\end{figure}

The graphical representations of the five different types of convolutions
given in Tab.~\ref{tab::conv} are displayed in Fig.~\ref{fig::sigxp}.

\begin{table}[t]
%  \centering
  \renewcommand{\arraystretch}{2.0}
  \begin{tabular}{c|c|c}  
    \hline
    convolution & order & file name
    \\
    \hline
    $(\tilde\sigma_{ij}^{(n)}/x) \otimes P_{jk}^{(r)}$
    &
    \begin{minipage}{8em}
      \centering
      N$^{(n+r)}$LO\\
      $n=0,1,2$;\\ $r=1,2,3$
    \end{minipage}
    &
    \begin{minipage}{17em}
      \small
      \verb|Stil<n><ij>-P<r><jk>.m|
    \end{minipage}
    \\
    \hline
    \begin{minipage}{10em}
      \centering
      $(\tilde\sigma_{ij}^{(n)}/x) \otimes P_{jk}^{(r)}\otimes P_{kl}^{(s)}$
      \\
      $P_{il}^{(s)} \otimes (\tilde\sigma_{ij}^{(n)}/x) \otimes P_{jk}^{(r)}$
    \end{minipage}
    &
    \begin{minipage}{8em}
      \centering
      N$^{(n+r+s)}$LO\\
      $n=0,1$;\\ $r,s=1,2$
    \end{minipage}
    &
    \begin{minipage}{17em}
      \small
      \verb|Stil<n><ij>-P<r><jk>-P<s><kl>.m|
      \\
      \verb|P<s><il>-Stil<n><ij>-P<r><jk>.m|
    \end{minipage}
    \\
    \hline
    \begin{minipage}{12em}
      \centering
      $(\tilde\sigma_{ij}^{(0)}/x) \otimes P_{jk}^{(1)}\otimes P_{kl}^{(1)}
      \otimes P_{lm}^{(1)}$
      \\
      $P_{im}^{(1)} \otimes (\tilde\sigma_{ij}^{(0)} /x)
      \otimes P_{jk}^{(1)}\otimes P_{kl}^{(1)}
      $
    \end{minipage}
    &
    \begin{minipage}{8em}
      \centering
      N$^3$LO
    \end{minipage}
    &
    \begin{minipage}{17em}
      \small
      \verb|Stil0<ij>-P1<jk>-P1<kl>-P1<lm>.m|
      \\
      \verb|P1<im>-Stil0<ij>-P1<jk>-P1<kl>.m|
    \end{minipage}
    \\
    \hline
  \end{tabular}
  \caption[]{\label{tab::conv}
    Classifications of the convolutions needed up to order N$^3$LO
    where $i,j,k,l,m \in \{g,q,\bar{q},q^\prime\}$. As far as splitting functions
    are concerned there is no difference between ``$q$'', ``$\bar{q}$\,'' and
    ``$q^\prime$\,''. Furthermore, we have 
    $\tilde\sigma^{(n)}_{ij} = \tilde\sigma^{(n)}_{ji}$.} 
\end{table}

Let us at the end list some convolutions which appear as building blocks
in the course of our calculation.
\begin{eqnarray}
  \lefteqn{\left[\left[\frac{\ln^3(1-x)}{1-x}\right]_+ 
      \otimes \left[\frac{1}{1-x}\right]_+\right]
    =
    \frac{6}{1-x}\left(
      \HPL{1, 1, 2}{x} + \HPL{1, 2, 1}{x} + \HPL{2, 1, 1}{x} 
      + \HPL{1, 1, 1, 0}{x}
    \right)
  }
  \nonumber\\&&\mbox{}
  + \frac{5}{4}      \left[\frac{\ln^4(1-x)}{1-x}\right]_+
  - \frac{\pi^2}{2}  \left[\frac{\ln^2(1-x)}{1-x}\right]_+
  + 6\zeta(3)        \left[\frac{\ln(1-x)}{1-x}\right]_+
  - \frac{\pi^4}{15} \left[\frac{1}{1-x}\right]_+
  \nonumber\\&&\mbox{}
  + 6\zeta(5)        \delta(1-x)
  \,,
  \nonumber\\
  \lefteqn{\left[\left[\frac{\ln^4(1-x)}{1-x}\right]_+ 
      \otimes \left[\frac{1}{1-x}\right]_+\right]
    =
    -\frac{24}{1-x}\left(
      \HPL{1, 1, 1, 2}{x} + \HPL{1, 1, 2, 1}{x} + \HPL{1, 2, 1, 1}{x} 
    \right.
  }
  \nonumber\\&&\left.\mbox{}
    + \HPL{2, 1, 1, 1}{x} 
    + \HPL{1, 1, 1, 1, 0}{x}
  \right)
  + \frac{6}{5}      \left[\frac{\ln^5(1-x)}{1-x}\right]_+
  - \frac{2\pi^2}{3} \left[\frac{\ln^3(1-x)}{1-x}\right]_+
  \nonumber\\&&\mbox{}
  + 12\zeta(3)       \left[\frac{\ln^2(1-x)}{1-x}\right]_+
  - \frac{4\pi^4}{15}\left[\frac{\ln(1-x)}{1-x}\right]_+
  + 24\zeta(5)       \left[\frac{1}{1-x}\right]_+
  - \frac{8\pi^6}{315}\delta(1-x)
  \,,
  \nonumber\\
  \lefteqn{\left[\HPL{1, 1, 1, 0}{x}
      \otimes \left[\frac{1}{1-x}\right]_+\right]
    =
  - \HPL{1, 1, 1, 2}{x} 
  - \HPL{1, 1, 2, 0}{x} 
  - \HPL{1, 2, 1, 0}{x} 
  - \HPL{2, 1, 1, 0}{x} 
  }
  \nonumber\\&&\mbox{}
  - \HPL{1, 1, 1, 0, 0}{x} 
  - 3\HPL{1, 1, 1, 1, 0}{x} 
  + \frac{\pi^2}{6}\HPL{1, 1, 1}{x}
  - 2\zeta(3)\HPL{1, 1}{x}
  - \frac{\pi^4}{72}\HPL{1}{x}
  \nonumber\\&&\mbox{}
  + \frac{\pi^2}{6}\zeta(3)
  - 3\zeta(5)
  \,.
\end{eqnarray}

%- }}}
%- {{{ Summary

\section{\label{sec::sum}Summary}

In this paper we computed the NNLO partonic cross sections for the
Higgs boson production to order $\epsilon$ which constitutes a building block
for a N$^3$LO calculation. Furthermore an algorithm has been developed which
allows the evaluation of all convolutions of a partonic cross section with
splitting functions needed at N$^3$LO. The described procedure might also be
interesting for other processes like Drell-Yan production.
We provide computer-readable results~\cite{progdata} 
both for the partonic cross sections
and for all the convolutions entering the N$^3$LO calculation of the partonic
cross section for Higgs boson production at the LHC.

%- }}}

%- {{{ Ackn.:

\section*{Acknowledgements}

This work was supported by the DFG through the SFB/TR~9 ``Computational
Particle Physics''.

%- }}}

\begin {appendix}

\section{\label{app::conv}Convolutions and names of result files}

Tables~\ref{tab::S} and~\ref{tab::P} list the convolutions 
entering our calculation together
with the names of the result files~\cite{progdata}. For convenience we use the
notation $\bar{q}\equiv b$ and $q^\prime \equiv p$.

\begin{table}
  \begin{center}
    {\scalefont{0.65}
      \begin{tabular}{ll||ll}
        \hline 
        & \\
$P_{gg}^{(1)} \otimes (\tilde\sigma_{gg}^{(0)}/x) \otimes P_{gg}^{(1)}$ & \verb|P1gg-Stil0gg-P1gg| & $P_{gg}^{(1)} \otimes (\tilde\sigma_{gg}^{(0)}/x) \otimes P_{gg}^{(1)} \otimes P_{gg}^{(1)}$ & \verb|P1gg-Stil0gg-P1gg-P1gg| \\ 
$P_{gg}^{(1)} \otimes (\tilde\sigma_{gg}^{(0)}/x) \otimes P_{gg}^{(1)} \otimes P_{gq}^{(1)}$ & \verb|P1gg-Stil0gg-P1gg-P1gq| & $P_{gg}^{(1)} \otimes (\tilde\sigma_{gg}^{(0)}/x) \otimes P_{gq}^{(1)}$ & \verb|P1gg-Stil0gg-P1gq| \\ 
$P_{gg}^{(1)} \otimes (\tilde\sigma_{gg}^{(0)}/x) \otimes P_{gq}^{(1)} \otimes P_{qg}^{(1)}$ & \verb|P1gg-Stil0gg-P1gq-P1qg| & $P_{gg}^{(1)} \otimes (\tilde\sigma_{gg}^{(0)}/x) \otimes P_{gq}^{(1)} \otimes P_{qq}^{(1)}$ & \verb|P1gg-Stil0gg-P1gq-P1qq| \\ 
$P_{gg}^{(1)} \otimes (\tilde\sigma_{gg}^{(0)}/x) \otimes P_{gg}^{(2)}$ & \verb|P1gg-Stil0gg-P2gg| & $P_{gg}^{(1)} \otimes (\tilde\sigma_{gg}^{(0)}/x) \otimes P_{gq}^{(2)}$ & \verb|P1gg-Stil0gg-P2gq| \\ 
$P_{gg}^{(1)} \otimes (\tilde\sigma_{gg}^{(1)}/x) \otimes P_{gg}^{(1)}$ & \verb|P1gg-Stil1gg-P1gg| & $P_{gg}^{(1)} \otimes (\tilde\sigma_{gg}^{(1)}/x) \otimes P_{gq}^{(1)}$ & \verb|P1gg-Stil1gg-P1gq| \\ 
$P_{gq}^{(1)} \otimes (\tilde\sigma_{gg}^{(0)}/x) \otimes P_{gg}^{(1)}$ & \verb|P1gq-Stil0gg-P1gg| & $P_{gq}^{(1)} \otimes (\tilde\sigma_{gg}^{(0)}/x) \otimes P_{gg}^{(1)} \otimes P_{gg}^{(1)}$ & \verb|P1gq-Stil0gg-P1gg-P1gg| \\ 
$P_{gq}^{(1)} \otimes (\tilde\sigma_{gg}^{(0)}/x) \otimes P_{gg}^{(1)} \otimes P_{gq}^{(1)}$ & \verb|P1gq-Stil0gg-P1gg-P1gq| & $P_{gq}^{(1)} \otimes (\tilde\sigma_{gg}^{(0)}/x) \otimes P_{gq}^{(1)}$ & \verb|P1gq-Stil0gg-P1gq| \\ 
$P_{gq}^{(1)} \otimes (\tilde\sigma_{gg}^{(0)}/x) \otimes P_{gq}^{(1)} \otimes P_{qg}^{(1)}$ & \verb|P1gq-Stil0gg-P1gq-P1qg| & $P_{gq}^{(1)} \otimes (\tilde\sigma_{gg}^{(0)}/x) \otimes P_{gq}^{(1)} \otimes P_{qq}^{(1)}$ & \verb|P1gq-Stil0gg-P1gq-P1qq| \\ 
$P_{gq}^{(1)} \otimes (\tilde\sigma_{gg}^{(0)}/x) \otimes P_{gg}^{(2)}$ & \verb|P1gq-Stil0gg-P2gg| & $P_{gq}^{(1)} \otimes (\tilde\sigma_{gg}^{(0)}/x) \otimes P_{gq}^{(2)}$ & \verb|P1gq-Stil0gg-P2gq| \\ 
$P_{gq}^{(1)} \otimes (\tilde\sigma_{gg}^{(1)}/x) \otimes P_{gg}^{(1)}$ & \verb|P1gq-Stil1gg-P1gg| & $P_{gq}^{(1)} \otimes (\tilde\sigma_{gg}^{(1)}/x) \otimes P_{gq}^{(1)}$ & \verb|P1gq-Stil1gg-P1gq| \\ 
$P_{qg}^{(1)} \otimes (\tilde\sigma_{qb}^{(1)}/x) \otimes P_{qg}^{(1)}$ & \verb|P1qg-Stil1qb-P1qg| & $P_{qg}^{(1)} \otimes (\tilde\sigma_{qb}^{(1)}/x) \otimes P_{qq}^{(1)}$ & \verb|P1qg-Stil1qb-P1qq| \\ 
$P_{qg}^{(1)} \otimes (\tilde\sigma_{qg}^{(1)}/x) \otimes P_{gg}^{(1)}$ & \verb|P1qg-Stil1qg-P1gg| & $P_{qg}^{(1)} \otimes (\tilde\sigma_{qg}^{(1)}/x) \otimes P_{gq}^{(1)}$ & \verb|P1qg-Stil1qg-P1gq| \\ 
$P_{qq}^{(1)} \otimes (\tilde\sigma_{qb}^{(1)}/x) \otimes P_{qg}^{(1)}$ & \verb|P1qq-Stil1qb-P1qg| & $P_{qq}^{(1)} \otimes (\tilde\sigma_{qb}^{(1)}/x) \otimes P_{qq}^{(1)}$ & \verb|P1qq-Stil1qb-P1qq| \\ 
$P_{qq}^{(1)} \otimes (\tilde\sigma_{qg}^{(1)}/x) \otimes P_{gg}^{(1)}$ & \verb|P1qq-Stil1qg-P1gg| & $P_{qq}^{(1)} \otimes (\tilde\sigma_{qg}^{(1)}/x) \otimes P_{gq}^{(1)}$ & \verb|P1qq-Stil1qg-P1gq| \\ 
$P_{gg}^{(2)} \otimes (\tilde\sigma_{gg}^{(0)}/x) \otimes P_{gg}^{(1)}$ & \verb|P2gg-Stil0gg-P1gg| & $P_{gg}^{(2)} \otimes (\tilde\sigma_{gg}^{(0)}/x) \otimes P_{gq}^{(1)}$ & \verb|P2gg-Stil0gg-P1gq| \\ 
$P_{gq}^{(2)} \otimes (\tilde\sigma_{gg}^{(0)}/x) \otimes P_{gg}^{(1)}$ & \verb|P2gq-Stil0gg-P1gg| & $P_{gq}^{(2)} \otimes (\tilde\sigma_{gg}^{(0)}/x) \otimes P_{gq}^{(1)}$ & \verb|P2gq-Stil0gg-P1gq| \\ 
$(\tilde\sigma_{gg}^{(0)}/x) \otimes P_{gg}^{(1)}$ & \verb|Stil0gg-P1gg| & $(\tilde\sigma_{gg}^{(0)}/x) \otimes P_{gg}^{(1)} \otimes P_{gg}^{(1)}$ & \verb|Stil0gg-P1gg-P1gg| \\ 
$(\tilde\sigma_{gg}^{(0)}/x) \otimes P_{gg}^{(1)} \otimes P_{gg}^{(1)} \otimes P_{gg}^{(1)}$ & \verb|Stil0gg-P1gg-P1gg-P1gg| & $(\tilde\sigma_{gg}^{(0)}/x) \otimes P_{gg}^{(1)} \otimes P_{gg}^{(1)} \otimes P_{gq}^{(1)}$ & \verb|Stil0gg-P1gg-P1gg-P1gq| \\ 
$(\tilde\sigma_{gg}^{(0)}/x) \otimes P_{gg}^{(1)} \otimes P_{gq}^{(1)}$ & \verb|Stil0gg-P1gg-P1gq| & $(\tilde\sigma_{gg}^{(0)}/x) \otimes P_{gg}^{(1)} \otimes P_{gq}^{(1)} \otimes P_{qg}^{(1)}$ & \verb|Stil0gg-P1gg-P1gq-P1qg| \\ 
$(\tilde\sigma_{gg}^{(0)}/x) \otimes P_{gg}^{(1)} \otimes P_{gq}^{(1)} \otimes P_{qq}^{(1)}$ & \verb|Stil0gg-P1gg-P1gq-P1qq| & $(\tilde\sigma_{gg}^{(0)}/x) \otimes P_{gg}^{(1)} \otimes P_{gg}^{(2)}$ & \verb|Stil0gg-P1gg-P2gg| \\ 
$(\tilde\sigma_{gg}^{(0)}/x) \otimes P_{gg}^{(1)} \otimes P_{gq}^{(2)}$ & \verb|Stil0gg-P1gg-P2gq| & $(\tilde\sigma_{gg}^{(0)}/x) \otimes P_{gq}^{(1)}$ & \verb|Stil0gg-P1gq| \\ 
$(\tilde\sigma_{gg}^{(0)}/x) \otimes P_{gq}^{(1)} \otimes P_{qg}^{(1)}$ & \verb|Stil0gg-P1gq-P1qg| & $(\tilde\sigma_{gg}^{(0)}/x) \otimes P_{gq}^{(1)} \otimes P_{qg}^{(1)} \otimes P_{gg}^{(1)}$ & \verb|Stil0gg-P1gq-P1qg-P1gg| \\ 
$(\tilde\sigma_{gg}^{(0)}/x) \otimes P_{gq}^{(1)} \otimes P_{qg}^{(1)} \otimes P_{gq}^{(1)}$ & \verb|Stil0gg-P1gq-P1qg-P1gq| & $(\tilde\sigma_{gg}^{(0)}/x) \otimes P_{gq}^{(1)} \otimes P_{qq}^{(1)}$ & \verb|Stil0gg-P1gq-P1qq| \\ 
$(\tilde\sigma_{gg}^{(0)}/x) \otimes P_{gq}^{(1)} \otimes P_{qq}^{(1)} \otimes P_{qg}^{(1)}$ & \verb|Stil0gg-P1gq-P1qq-P1qg| & $(\tilde\sigma_{gg}^{(0)}/x) \otimes P_{gq}^{(1)} \otimes P_{qq}^{(1)} \otimes P_{qq}^{(1)}$ & \verb|Stil0gg-P1gq-P1qq-P1qq| \\ 
$(\tilde\sigma_{gg}^{(0)}/x) \otimes P_{gq}^{(1)} \otimes P_{qg}^{(2)}$ & \verb|Stil0gg-P1gq-P2qg| & $(\tilde\sigma_{gg}^{(0)}/x) \otimes P_{gq}^{(1)} \otimes P_{qq}^{(2)}$ & \verb|Stil0gg-P1gq-P2qq| \\ 
$(\tilde\sigma_{gg}^{(0)}/x) \otimes P_{gg}^{(2)}$ & \verb|Stil0gg-P2gg| & $(\tilde\sigma_{gg}^{(0)}/x) \otimes P_{gg}^{(2)} \otimes P_{gg}^{(1)}$ & \verb|Stil0gg-P2gg-P1gg| \\ 
$(\tilde\sigma_{gg}^{(0)}/x) \otimes P_{gg}^{(2)} \otimes P_{gq}^{(1)}$ & \verb|Stil0gg-P2gg-P1gq| & $(\tilde\sigma_{gg}^{(0)}/x) \otimes P_{gq}^{(2)}$ & \verb|Stil0gg-P2gq| \\ 
$(\tilde\sigma_{gg}^{(0)}/x) \otimes P_{gq}^{(2)} \otimes P_{qg}^{(1)}$ & \verb|Stil0gg-P2gq-P1qg| & $(\tilde\sigma_{gg}^{(0)}/x) \otimes P_{gq}^{(2)} \otimes P_{qq}^{(1)}$ & \verb|Stil0gg-P2gq-P1qq| \\ 
$(\tilde\sigma_{gg}^{(0)}/x) \otimes P_{gg}^{(3)}$ & \verb|Stil0gg-P3gg| & $(\tilde\sigma_{gg}^{(0)}/x) \otimes P_{gq}^{(3)}$ & \verb|Stil0gg-P3gq| \\ 
$(\tilde\sigma_{gg}^{(1)}/x) \otimes P_{gg}^{(1)}$ & \verb|Stil1gg-P1gg| & $(\tilde\sigma_{gg}^{(1)}/x) \otimes P_{gg}^{(1)} \otimes P_{gg}^{(1)}$ & \verb|Stil1gg-P1gg-P1gg| \\ 
$(\tilde\sigma_{gg}^{(1)}/x) \otimes P_{gg}^{(1)} \otimes P_{gq}^{(1)}$ & \verb|Stil1gg-P1gg-P1gq| & $(\tilde\sigma_{gg}^{(1)}/x) \otimes P_{gq}^{(1)}$ & \verb|Stil1gg-P1gq| \\ 
$(\tilde\sigma_{gg}^{(1)}/x) \otimes P_{gq}^{(1)} \otimes P_{qg}^{(1)}$ & \verb|Stil1gg-P1gq-P1qg| & $(\tilde\sigma_{gg}^{(1)}/x) \otimes P_{gq}^{(1)} \otimes P_{qq}^{(1)}$ & \verb|Stil1gg-P1gq-P1qq| \\ 
$(\tilde\sigma_{gg}^{(1)}/x) \otimes P_{gg}^{(2)}$ & \verb|Stil1gg-P2gg| & $(\tilde\sigma_{gg}^{(1)}/x) \otimes P_{gq}^{(2)}$ & \verb|Stil1gg-P2gq| \\ 
$(\tilde\sigma_{qb}^{(1)}/x) \otimes P_{qg}^{(1)}$ & \verb|Stil1qb-P1qg| & $(\tilde\sigma_{qb}^{(1)}/x) \otimes P_{qg}^{(1)} \otimes P_{gg}^{(1)}$ & \verb|Stil1qb-P1qg-P1gg| \\ 
$(\tilde\sigma_{qb}^{(1)}/x) \otimes P_{qg}^{(1)} \otimes P_{gq}^{(1)}$ & \verb|Stil1qb-P1qg-P1gq| & $(\tilde\sigma_{qb}^{(1)}/x) \otimes P_{qq}^{(1)}$ & \verb|Stil1qb-P1qq| \\ 
$(\tilde\sigma_{qb}^{(1)}/x) \otimes P_{qq}^{(1)} \otimes P_{qg}^{(1)}$ & \verb|Stil1qb-P1qq-P1qg| & $(\tilde\sigma_{qb}^{(1)}/x) \otimes P_{qq}^{(1)} \otimes P_{qq}^{(1)}$ & \verb|Stil1qb-P1qq-P1qq| \\ 
$(\tilde\sigma_{qb}^{(1)}/x) \otimes P_{qg}^{(2)}$ & \verb|Stil1qb-P2qg| & $(\tilde\sigma_{qb}^{(1)}/x) \otimes P_{qq}^{(2)}$ & \verb|Stil1qb-P2qq| \\ 
$(\tilde\sigma_{qg}^{(1)}/x) \otimes P_{gg}^{(1)}$ & \verb|Stil1qg-P1gg| & $(\tilde\sigma_{qg}^{(1)}/x) \otimes P_{gg}^{(1)} \otimes P_{gg}^{(1)}$ & \verb|Stil1qg-P1gg-P1gg| \\ 
$(\tilde\sigma_{qg}^{(1)}/x) \otimes P_{gg}^{(1)} \otimes P_{gq}^{(1)}$ & \verb|Stil1qg-P1gg-P1gq| & $(\tilde\sigma_{qg}^{(1)}/x) \otimes P_{gq}^{(1)}$ & \verb|Stil1qg-P1gq| \\ 
$(\tilde\sigma_{qg}^{(1)}/x) \otimes P_{gq}^{(1)} \otimes P_{qg}^{(1)}$ & \verb|Stil1qg-P1gq-P1qg| & $(\tilde\sigma_{qg}^{(1)}/x) \otimes P_{gq}^{(1)} \otimes P_{qq}^{(1)}$ & \verb|Stil1qg-P1gq-P1qq| \\ 
$(\tilde\sigma_{qg}^{(1)}/x) \otimes P_{qg}^{(1)}$ & \verb|Stil1qg-P1qg| & $(\tilde\sigma_{qg}^{(1)}/x) \otimes P_{qg}^{(1)} \otimes P_{gg}^{(1)}$ & \verb|Stil1qg-P1qg-P1gg| \\ 
$(\tilde\sigma_{qg}^{(1)}/x) \otimes P_{qg}^{(1)} \otimes P_{gq}^{(1)}$ & \verb|Stil1qg-P1qg-P1gq| & $(\tilde\sigma_{qg}^{(1)}/x) \otimes P_{qq}^{(1)}$ & \verb|Stil1qg-P1qq| \\ 
$(\tilde\sigma_{qg}^{(1)}/x) \otimes P_{qq}^{(1)} \otimes P_{qg}^{(1)}$ & \verb|Stil1qg-P1qq-P1qg| & $(\tilde\sigma_{qg}^{(1)}/x) \otimes P_{qq}^{(1)} \otimes P_{qq}^{(1)}$ & \verb|Stil1qg-P1qq-P1qq| \\ 
$(\tilde\sigma_{qg}^{(1)}/x) \otimes P_{gg}^{(2)}$ & \verb|Stil1qg-P2gg| & $(\tilde\sigma_{qg}^{(1)}/x) \otimes P_{gq}^{(2)}$ & \verb|Stil1qg-P2gq| \\ 
$(\tilde\sigma_{qg}^{(1)}/x) \otimes P_{qg}^{(2)}$ & \verb|Stil1qg-P2qg| & $(\tilde\sigma_{qg}^{(1)}/x) \otimes P_{qq}^{(2)}$ & \verb|Stil1qg-P2qq| \\ 
$(\tilde\sigma_{gg}^{(2)}/x) \otimes P_{gg}^{(1)}$ & \verb|Stil2gg-P1gg| & $(\tilde\sigma_{gg}^{(2)}/x) \otimes P_{gq}^{(1)}$ & \verb|Stil2gg-P1gq| \\ 
$(\tilde\sigma_{qb}^{(2)}/x) \otimes P_{qg}^{(1)}$ & \verb|Stil2qb-P1qg| & $(\tilde\sigma_{qb}^{(2)}/x) \otimes P_{qq}^{(1)}$ & \verb|Stil2qb-P1qq| \\ 
$(\tilde\sigma_{qg}^{(2)}/x) \otimes P_{gg}^{(1)}$ & \verb|Stil2qg-P1gg| & $(\tilde\sigma_{qg}^{(2)}/x) \otimes P_{gq}^{(1)}$ & \verb|Stil2qg-P1gq| \\ 
$(\tilde\sigma_{qg}^{(2)}/x) \otimes P_{qg}^{(1)}$ & \verb|Stil2qg-P1qg| & $(\tilde\sigma_{qg}^{(2)}/x) \otimes P_{qq}^{(1)}$ & \verb|Stil2qg-P1qq| \\ 
$(\tilde\sigma_{qp}^{(2)}/x) \otimes P_{qg}^{(1)}$ & \verb|Stil2qp-P1qg| & $(\tilde\sigma_{qp}^{(2)}/x) \otimes P_{qq}^{(1)}$ & \verb|Stil2qp-P1qq| \\ 
$(\tilde\sigma_{qq}^{(2)}/x) \otimes P_{qg}^{(1)}$ & \verb|Stil2qq-P1qg| & $(\tilde\sigma_{qq}^{(2)}/x) \otimes P_{qq}^{(1)}$ & \verb|Stil2qq-P1qq| \\ 
        & \\
        \hline
      \end{tabular}
    }
  \end{center}
  \caption{\label{tab::S}Correlation of convolutions involving
    the partonic cross section and names of result file.}
\end{table}

\begin{table}
  \begin{center}
    {\scalefont{0.7}
      \begin{tabular}{ll||ll}
        \hline
        & \\
$P_{gg}^{(1)} \otimes P_{gg}^{(1)}$ & \verb|P1gg-P1gg| & $P_{gg}^{(1)} \otimes P_{gg}^{(1)} \otimes P_{gg}^{(1)}$ & \verb|P1gg-P1gg-P1gg| \\ 
$P_{gg}^{(1)} \otimes P_{gg}^{(1)} \otimes P_{gq}^{(1)}$ & \verb|P1gg-P1gg-P1gq| & $P_{gg}^{(1)} \otimes P_{gq}^{(1)}$ & \verb|P1gg-P1gq| \\ 
$P_{gg}^{(1)} \otimes P_{gq}^{(1)} \otimes P_{qg}^{(1)}$ & \verb|P1gg-P1gq-P1qg| & $P_{gg}^{(1)} \otimes P_{gq}^{(1)} \otimes P_{qq}^{(1)}$ & \verb|P1gg-P1gq-P1qq| \\ 
$P_{gg}^{(1)} \otimes P_{gg}^{(2)}$ & \verb|P1gg-P2gg| & $P_{gg}^{(1)} \otimes P_{gq}^{(2)}$ & \verb|P1gg-P2gq| \\ 
$P_{gq}^{(1)} \otimes P_{gg}^{(1)}$ & \verb|P1gq-P1gg| & $P_{gq}^{(1)} \otimes P_{gg}^{(1)} \otimes P_{gg}^{(1)}$ & \verb|P1gq-P1gg-P1gg| \\ 
$P_{gq}^{(1)} \otimes P_{gg}^{(1)} \otimes P_{gq}^{(1)}$ & \verb|P1gq-P1gg-P1gq| & $P_{gq}^{(1)} \otimes P_{gq}^{(1)}$ & \verb|P1gq-P1gq| \\ 
$P_{gq}^{(1)} \otimes P_{gq}^{(1)} \otimes P_{qg}^{(1)}$ & \verb|P1gq-P1gq-P1qg| & $P_{gq}^{(1)} \otimes P_{gq}^{(1)} \otimes P_{qq}^{(1)}$ & \verb|P1gq-P1gq-P1qq| \\ 
$P_{gq}^{(1)} \otimes P_{qg}^{(1)}$ & \verb|P1gq-P1qg| & $P_{gq}^{(1)} \otimes P_{qg}^{(1)} \otimes P_{gg}^{(1)}$ & \verb|P1gq-P1qg-P1gg| \\ 
$P_{gq}^{(1)} \otimes P_{qg}^{(1)} \otimes P_{gq}^{(1)}$ & \verb|P1gq-P1qg-P1gq| & $P_{gq}^{(1)} \otimes P_{qq}^{(1)}$ & \verb|P1gq-P1qq| \\ 
$P_{gq}^{(1)} \otimes P_{qq}^{(1)} \otimes P_{qg}^{(1)}$ & \verb|P1gq-P1qq-P1qg| & $P_{gq}^{(1)} \otimes P_{qq}^{(1)} \otimes P_{qq}^{(1)}$ & \verb|P1gq-P1qq-P1qq| \\ 
$P_{gq}^{(1)} \otimes P_{gg}^{(2)}$ & \verb|P1gq-P2gg| & $P_{gq}^{(1)} \otimes P_{gq}^{(2)}$ & \verb|P1gq-P2gq| \\ 
$P_{gq}^{(1)} \otimes P_{qg}^{(2)}$ & \verb|P1gq-P2qg| & $P_{gq}^{(1)} \otimes P_{qq}^{(2)}$ & \verb|P1gq-P2qq| \\ 
$P_{qg}^{(1)} \otimes P_{gg}^{(1)}$ & \verb|P1qg-P1gg| & $P_{qg}^{(1)} \otimes P_{gq}^{(1)}$ & \verb|P1qg-P1gq| \\ 
$P_{qg}^{(1)} \otimes P_{qg}^{(1)}$ & \verb|P1qg-P1qg| & $P_{qq}^{(1)} \otimes P_{gg}^{(1)}$ & \verb|P1qq-P1gg| \\ 
$P_{qq}^{(1)} \otimes P_{gq}^{(1)}$ & \verb|P1qq-P1gq| & $P_{qq}^{(1)} \otimes P_{qg}^{(1)}$ & \verb|P1qq-P1qg| \\ 
$P_{qq}^{(1)} \otimes P_{qq}^{(1)}$ & \verb|P1qq-P1qq| & $P_{gg}^{(2)} \otimes P_{gg}^{(1)}$ & \verb|P2gg-P1gg| \\ 
$P_{gg}^{(2)} \otimes P_{gq}^{(1)}$ & \verb|P2gg-P1gq| & $P_{gq}^{(2)} \otimes P_{gg}^{(1)}$ & \verb|P2gq-P1gg| \\ 
$P_{gq}^{(2)} \otimes P_{gq}^{(1)}$ & \verb|P2gq-P1gq| & $P_{gq}^{(2)} \otimes P_{qg}^{(1)}$ & \verb|P2gq-P1qg| \\ 
$P_{gq}^{(2)} \otimes P_{qq}^{(1)}$ & \verb|P2gq-P1qq| & $$ & \verb|| \\ 
        & \\
        \hline
      \end{tabular}
    }
    \caption{\label{tab::P}Correlation of convolutions involving only
      splitting functions and names of result file.}
  \end{center}
\end{table}
  
\end{appendix}

%- {{{ bibliography

%- }}}

\end{document}